# Large Enhancement of Electronic Thermal Conductivity and Lorenz Number in Topological Insulator $Bi_2Te_2Se$ Thin Films

Zhe Luo[1,#], Jifa Tian[2,#], Mithun Srinivasan[1], Yong P. Chen[2,3,]* and Xianfan Xu[1,]*

[1] *School of Mechanical Engineering and Birck Nanotechnology Center, Purdue University, West Lafayette, Indiana 47907, USA*

[2] *Department of Physics & Astronomy and Birck Nanotechnology Center, Purdue University, West Lafayette, Indiana 47907, USA*

[3] *School of Electrical and Computer Engineering, Purdue University, West Lafayette, Indiana 47907, USA*

[#] These authors contributed equally to this work.

* To whom correspondence should be addressed. Email: yongchen@purdue.edu, xxu@ecn.purdue.edu

ABSTRACT

Topological insulators (TI) have attracted extensive research effort due to their insulating bulk states but conducting surface states. However, investigation and understanding of thermal transport in topological insulators, particularly the effect of surface states are lacking. In this work, we studied thickness-dependent in-plane thermal conductivity of $Bi_2Te_2Se$ TI thin films. A large enhancement of both thermal and electrical conductivity was observed for films with thicknesses below 20 nm, which is attributed to the surface states and bulk-insulating nature of these films. Surprisingly, a surface Lorenz number of over 10 times the Sommerfeld value was found. Transport measurements indicated that the surface is near the charge neutrality point when the film thickness is below 20 nm. Possible reasons for the large Lorenz number include the electrical and thermal current decoupling in the surface state Dirac fluid and the bipolar diffusion transport involving surface states.

MAIN TEXT

A three-dimensional (3D) topological insulator (TI) has a bulk band gap and behaves as an insulator in its interior, but possesses protected conducting electronic states on its surface known as topological surface states (TSS)[1–3]. With the unique spin texture and linear energy-momentum dispersion of the TSS, TIs exhibit exotic quantum-physical properties and have been extensively studied for the past decade as a novel class of materials [4–9]. From a practical perspective, the spin-momentum locking of TSS may yield a dissipationless spin current at the TI surface, which is of great potential in applications such as spintronics and quantum computation[1,2]. Despite the abundant research effort in TIs, studies regarding thermal transport in TSS are rare. In usual 3D semiconductor solids, thermal energy is transferred through bulk electrons (or holes) and phonons. For TIs, TSS provide yet another heat conducting channel involving backscattering-free, massless Dirac electrons, which may give rise to intriguing heat transfer phenomena. Among the TI materials, $Bi_2Te_2Se$ (BTS221) has been shown to have a minimal bulk contribution to electronic transport and prominent TSS properties[10–12], which serves as a good platform to study thermal transport arising from TSS carriers. Moreover, it has also been proposed that TIs could exhibit large thermoelectric figure-of-merit enhancement by proper manipulations of TSS[13–16], which brings more interest to the investigation of thermal transport of TI surface states.

Here we report studies of in-plane thermal conductivity of BTS221 TI films using Raman thermometry[17–22], supplemented by standard transport measurements of electrical conductivity. Raman thermometry is a well-established and validated technique that utilizes laser heating and measures the resulting temperature rise from Raman scattering spectra, and then extract thermal transport properties from temperature data. Our results and analysis show that both effective 3D electrical and thermal conductivity increase significantly as the BTS221 film thickness decreases, as a result of emerging contribution from the surface states which accounts for ~70% for electrical conductance and nearly half of the total thermal conductance when the thickness is below 20 nm. An exceptionally large Lorenz number of over 10 times of the Sommerfeld value is found, demonstrating that surface states provide a superior thermal transport channel compared with bulk electrons and phonons.

BTS221 thin films were prepared via tape-exfoliation from bulk crystals synthesized by the Bridgman technique. Such films have been shown to possess clear TSS transport properties at low temperature[23]. The exfoliated thin films were subsequently suspended on holey SiN membranes using a wet transfer technique for Raman thermometry measurements (see Methods and previous work[21,22]). To unveil the role of surface states in the overall thermal transport in BTS221, extensive effort was put forward to bring the thickness down to 8 nm in order to reduce the bulk contribution as much as possible. Figure 1a shows Raman spectra and an optical image of a suspended 10-nm flake at elevated temperature for calibration purpose. The $E_g^2$ mode was used as the Raman thermometer for its good sensitivity to temperature change, with the temperature coefficient $\chi_{Eg2} = -0.0113$ cm$^{-1}$/K where $\chi = \Delta\omega/\Delta T$, $\omega$ being the frequency and $T$ the temperature (Fig. 1b).

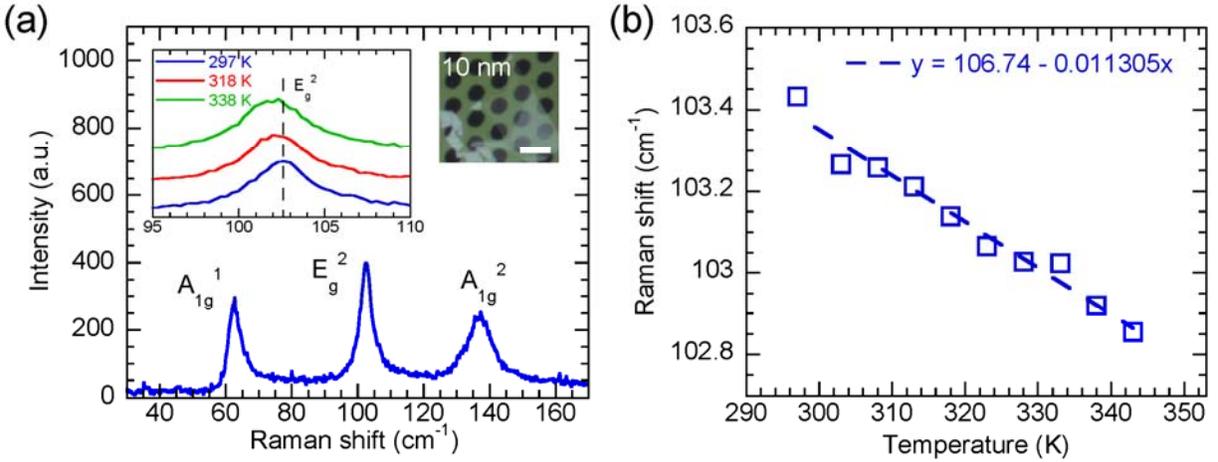

**Figure 1 | Raman thermometer calibration of a 10-nm thick BTS221 film. (a)** Raman spectrum of the film. Left inset: typical Raman spectra of the $E_g^2$ mode at various temperatures. Right inset: optical image of the transferred film, scale bar: 5 μm. **(b)** $E_g^2$ mode Raman shift vs. temperature. Dashed line is linear fit, whose slope is the temperature coefficient.

In the Raman thermometry measurements, the suspended BTS221 films were heated by a He-Ne laser at various power. The temperature rise from laser heating was determined by the shift in the $E_g^2$ Raman peak, and subsequently used in a 2D numerical heat transfer model to extract the in-plane thermal conductivity (see Methods and Ref. 22 for details). The measurements were carried out near room temperature (300 K) and the results are shown in Fig. 2a. As the thicknesses of the BTS221 films decrease to less than 20 nm, the measured effective in-plane thermal conductivity $k$ undergoes a rapid increase from ~1 W/mK to ~ 3 W/mK for films less than 10-nm thick, indicating the possible contributions from surface carriers. In addition, electrical transport measurements were conducted on a number of BTS221 devices. For these electrical transport measurements, BTS221 flakes were exfoliated to Si wafer with 300-nm thermally grown $SiO_2$ on top. Four Cr/Au electrodes were patterned using electron beam lithography (EBL) and standard lift-off processes, and four-probe sheet resistance measurements were carried out (Fig. 2b). Figure 2c shows the thickness-dependent effective 3D electrical conductivity $\sigma = 1/(R_S t)$ measured at room temperature where $R_S$ is the measured sheet resistance and $t$ is the film thickness. As seen in Fig. 2c, $\sigma$ also has an uprising trend when the thickness is less than ~20 nm, similar to the thermal conductivity results in Fig. 2a.

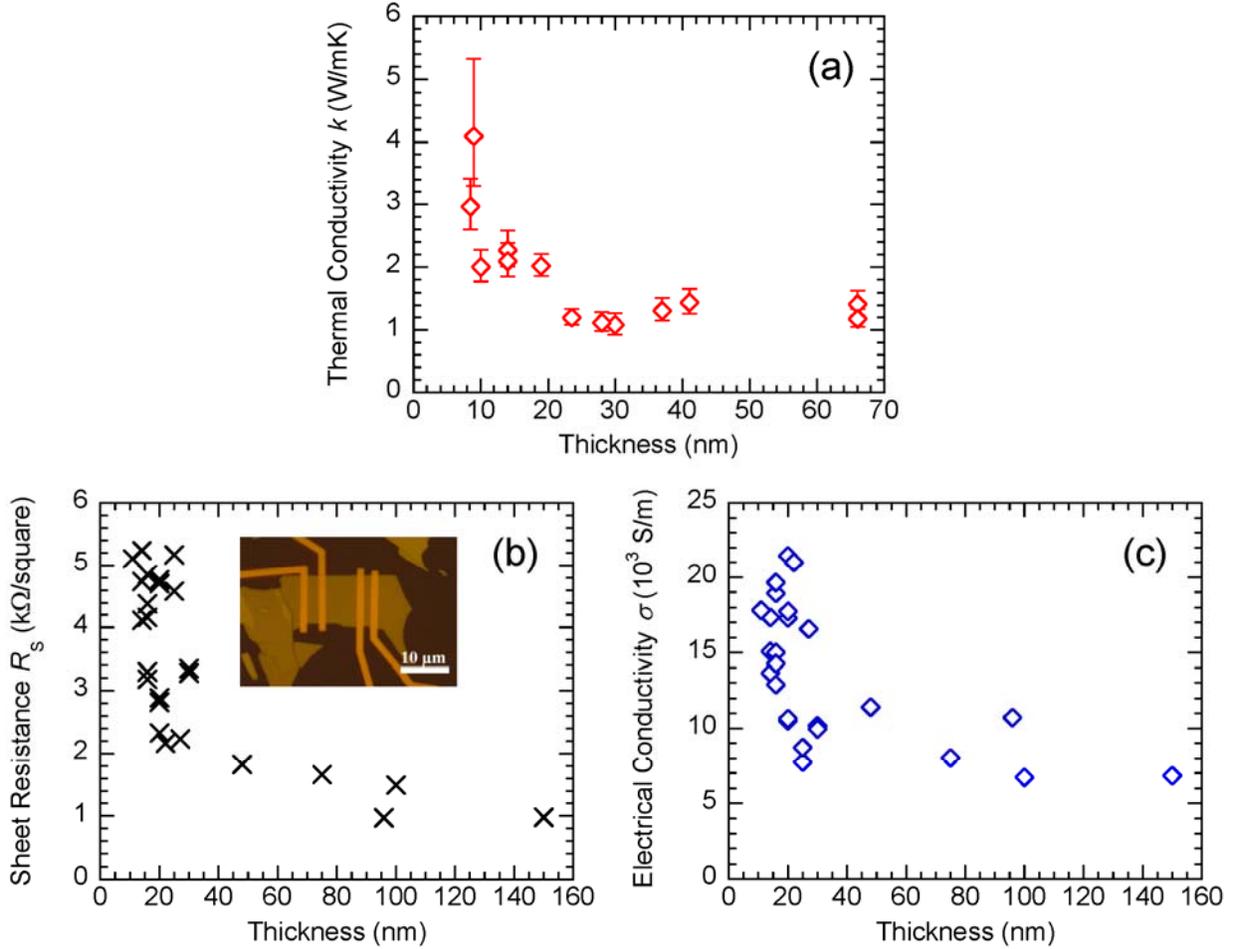

**Figure 2 | Thermal and electrical transport of BTS221 thin films. (a)** The thickness dependence of the effective in-plane thermal conductivity $k$ measured by the Raman thermometry. **(b)** The thickness dependence of the four-probe electrical sheet resistance $R_S$ measured in BTS221 devices. Inset: optical graph of a typical device. **(c)** Effective 3D electrical conductivity $\sigma$ calculated from (b). The relative uncertainty of data in (b) and (c) is ±2.5%. All measurements are made at room temperature.

The higher effective 3D electrical conductivity at smaller thicknesses in BTS221 films can be attributed to the increase of surface contribution due to the reduced bulk conduction.[17] Generally in TIs, the topological surface state gives a metallic conducting layer at the surface of the TI, while the bulk interior can be much less conducting if its carrier density (doping) is low. Furthermore, the surface state electrons can have high mobility[10,23] and weak electron-phonon coupling with the insulating bulk [24,25]. Our data show both high effective 3D electrical conductivity and high effective thermal conductivity when thickness decreases, possibly indicating increased contributions of surface states to both electrical and thermal transport. As such, we analyze the electrical conductivity and thermal conductivity data using a two-layer transport model accounting for the contributions from surface states and bulk. The total electrical conductance ($G$) and thermal conductance ($\Gamma$) are written as:

$$G = G_{surf} + G_{bulk} = G_{surf} + \sigma_{bulk} t \tag{1a}$$

$$\Gamma = \Gamma_{surf} + \Gamma_{bulk} = \Gamma_{surf} + k_{bulk} t \tag{1b}$$

The subscripts surf and bulk represent contributions from surface and bulk, respectively. $G = \sigma t$ and $\Gamma = kt$ are conductance for the entire film. Here underlies an assumption that the bulk thermal and electrical conductivity $k_{bulk}$ and $\sigma_{bulk}$ are independent of the thickness, which does not hold if the thickness is smaller than the mean-free-path of the heat and charge carriers. In this case a more complete model would be desired, but is beyond the scope here. However, if the reduction of bulk electrical and thermal conductivity due to smaller mean-free-path in thin films is considered, an even higher surface contribution will be obtained, which will be consistent with our main conclusions discussed below.

The thickness-dependent total electrical and thermal conductance calculated from the experimental data ($\sigma$ and $k$) are shown in Figs. 3a and b. The linear fit of both data show a non-zero intercept, which is attributed to the surface conductance, and the slope is the bulk conductivity: $G_{surf} = 0.16$ k$\Omega^{-1}$, $\sigma_{bulk} = 6.3 \times 10^3$ S/m, $\Gamma_{surf} = 1.4 \times 10^{-8}$ W/K, and $k_{bulk} = 1.0$ W/mK. Figure 3c shows the surface electrical (thermal) conductance to the total electrical (thermal) conductance ratio, $G_{surf}/G$ ($\Gamma_{surf}/\Gamma$). As the thickness decreases, the contribution from the surface conductance rises up from less than 20% up to 80% for electrical conductance, and up to 70% for thermal conductance. The extracted room-temperature bulk electrical resistivity $\rho = 1/\sigma \sim 0.016$ $\Omega\cdot$cm, is comparable with the reported resistivity 0.01–0.02 $\Omega\cdot$cm measured in thicker BTS221 bulk crystals[10]. The bulk thermal conductivity value is also consistent with previously reported theoretical and experimental data of $Bi_2Te_3$[26–28], which is of similar lattice structure as BTS221. The results show that, at sufficiently small thickness, the electrical and thermal transport of thin BTS221 film could be dominated by the surface even at room temperature.

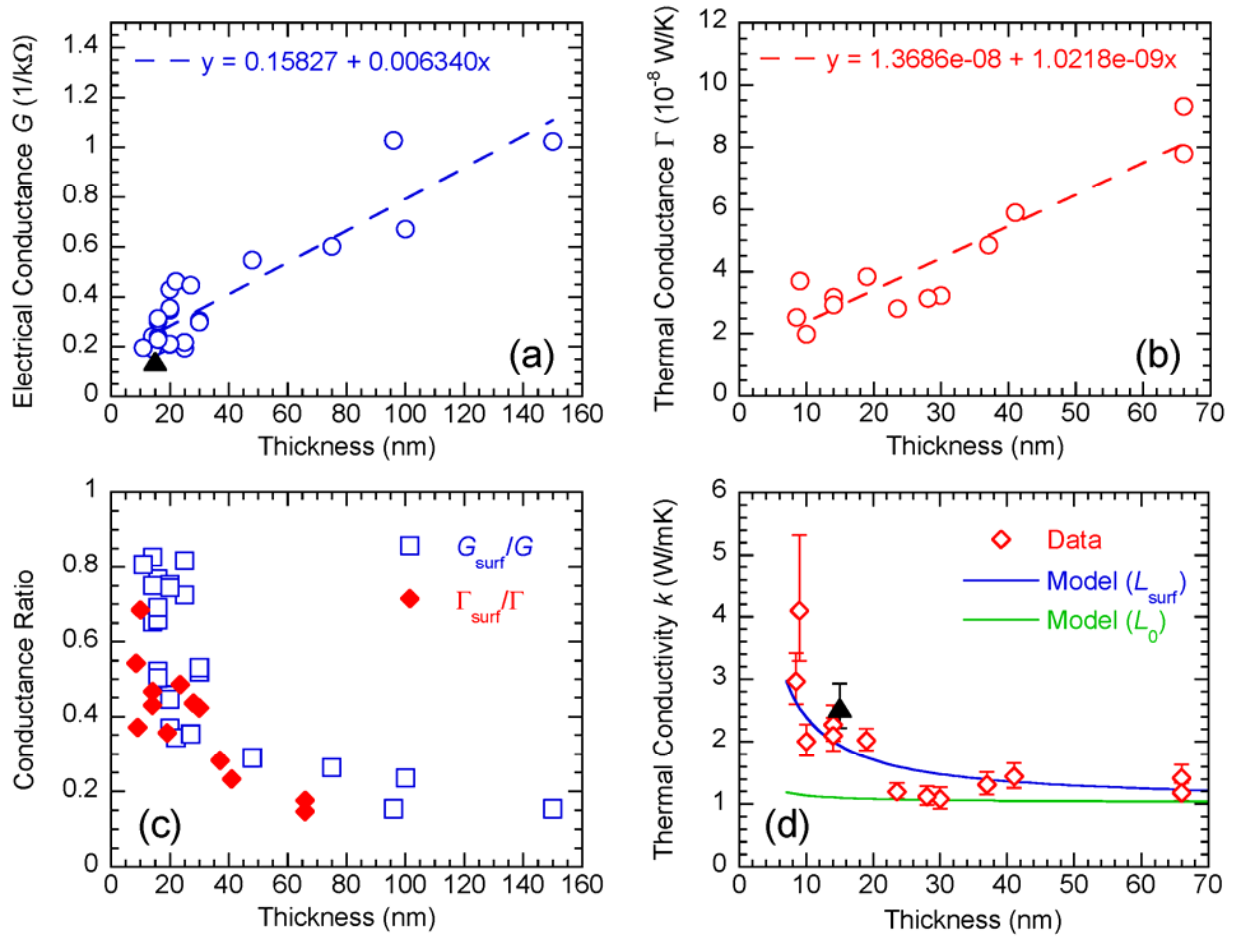

**Figure 3 | Analyses of the transport properties of BTS221 thin films to extract surface and bulk contributions.** (a) Electrical conductance $G_{tot}$ and a linear fit (dashed line). (b) Thermal conductance $\Gamma$ and linear fit. (c) Calculated ratio of surface electrical (thermal) conductance $G_{surf}$ ($\Gamma_{surf}$) to the total conductance $G$ ($\Gamma$). (d) Fitting the in-plane thermal conductivity $k$ (experimental

data from Fig. 2a) using the extracted Lorenz number $L$ and Sommerfeld value $L_0$ in Eq.(2). The black pyramid symbol in (a) and (d) represents the value obtained from a separate four-terminal BTS221-SiN device.

The relation between the thermal and electrical conductivity is described by the Wiedemann-Franz law, $k_e/\sigma = LT$, where $L$ is the Lorenz number. Rewriting it in conductance form gives $\Gamma/G = LT$. For the surface contribution to conductance in the measured BTS221 thin films, a surface Lorenz number $L_{surf}$ is defined here as $\Gamma_{surf}/G_{surf} = L_{surf}T$. Surprisingly, it is found that $L_{surf} = (2.9 \pm 1.0) \times 10^{-7}$ V$^2$/K$^2$, over 10 times larger than the Sommerfeld value $L_0 = 2.44 \times 10^{-8}$ V$^2$/K$^2$. To compare the experimentally measured total thermal conductivity with the model prediction using either $L_{surf}$ or $L_0$, the in-plane thermal conductivity is calculated using the following equation following Equation (1):

$$k = \frac{\Gamma_{surf}}{t} + k_{bulk} = \frac{LTG_{surf}}{t} + k_{bulk} \qquad (2)$$

We use both $L_{surf}$ and the Sommerfeld value $L_0$ in the calculation, and the calculated $k$ values are shown as solid lines in Fig. 3d. The line calculated using $L_{surf}$ fits the experimental data reasonably well, while the line using $L_0$ cannot capture the significant thermal conductivity rise for thinner films. Note that we are not able to separate the contributions of bulk electrons and phonons in $G$ and $\Gamma$, so the Lorenz number $L_{bulk}$ for bulk electrons cannot be extracted. Instead, we have used the commonly accepted Sommerfeld value $L_0$ to calculate the bulk electronic thermal conductivity which is presented below.

We further calculate the contributions from surface ($\Gamma_{surf}$), bulk electrons (using $L_0$ and fitted $\sigma_{bulk}$), and phonons (subtract the above two from measured $\Gamma$) to the total thermal conductance $\Gamma$, which is shown in Fig. 4. For the thickest film tested, surface state accounts for less than 20% of the total thermal conductance, and its contribution increases as the film thickness is reduced. For the thinner films whose thickness ranges from 8 to 30 nm, the surface state contribution can vary from ~40% up to 70%. In contrary, bulk electrons only account for ~5% of the thermal conduction. Therefore, in the BTS221 TI films, surface electrons and bulk phonons are the major heat carriers.

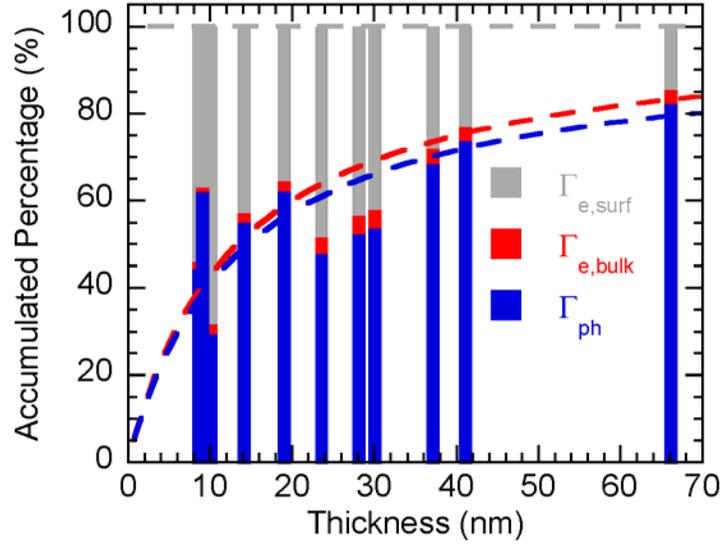

**Figure 4 | Percentage contribution to the total thermal conductance.** Contributions from bulk electrons and (bulk) phonons are calculated using $L_{TSS}$ for surface electrons and $L_0$ for bulk electrons. Dashed lines are predictions from the model.

It remains a concern that the $SiO_2$ dielectric layer underneath the TI film during the electrical measurements could alter the electronic properties of the surface states, for example by shifting the fermi level relative to the Dirac point[29,30]. There is also a concern that the BTS221 device fabrication process for the electrical measurements could alter the BTS221 films. To check this, a four-probe device patterned on SiN membrane with suspended BTS221 was prepared, and the same sample was used to measure both thermal and electrical conductivity at room temperature. The results are shown as the black pyramid symbol in Fig. 3a and d. It is seen that the measured electrical conductance and thermal conductivity are in good agreement with other experimental data, and they also lie in vicinity of the model prediction using the large $L_{surf}$.

Now we discuss the exceptionally large surface Lorenz number, which was derived from thermal and electrical conductivity data obtained from suspended samples for $k$ measurements and supported BTS221 devices on SiO$_2$/Si substrate for $\sigma$ measurements. The observed large surface Lorenz number $L_{\text{surf}}$ can be attributed to several possible origins. One is the effect of Dirac fluid of TSS electrons, with a strongly interacting electron-hole plasma formed near the Dirac point[31]. The Dirac fluid theory suggests that in the temperature range where the electron-electron scattering is the dominating charge carrier scattering mechanism, the hydrodynamic nature of the charge carriers leads to the decoupling of charge and (electronic) heat current. In this case, the charge current is limited by the viscosity of Dirac fluid, while the heat current carried by electrons (holes) is limited only by impurity scattering. This will inevitably lead to the breakdown of the Wiedemann-Franz law, which has been theoretically predicted in graphene[32,33]. These theoretical works show that, in a clean limit of graphene where the impurity scattering is negligible compared with electron-electron scattering, and in the temperature range $k_B T > \mu$ (chemical potential), i.e. when both electrons and holes are thermally excited, $L$ diverges to infinity. Indeed, a nearly 20-fold increase of the Lorenz number over the Sommerfeld value was recently experimentally observed in graphene near 75 K[31]. In our case, we find that, as the sample thickness decreases from 80 nm to below 20 nm, the extracted carrier density decreases more than one order of magnitude and the corresponding Fermi level becomes much closer to the Dirac point of TSS (Fig. S3). Furthermore, the observed nonlinearity of the Hall resistance in the samples indicates the coexistence of electrons and holes (Fig. S3). Thus, in these thin samples, the majority of electrical and/or thermal transport originates from both the electrons and holes near the Dirac point of TSS, where the coupling between electronic and heat currents due to the charge carrier (electron or hole) may be significantly weakened, resulting in the observed large Lorenz number.

Another possible reason for the observed large $L_{surf}$ is the bipolar diffusion in semiconductor materials[34]. Wiedemann-Franz law is based on the single band picture, so that when two or more bands contribute to transport, Wiedemann-Franz law can break down[35]. When both electrons and holes are thermally excited (for example, in non-degenerate semiconductors with small band gap at room or elevated temperature) and a temperature difference is created, the carriers diffuse to the colder region. In addition to the sum of the thermal transport by electrons and holes separately, the carrier recombination at the colder side releases energy, which enhances the total thermal conductivity[28,36]. This can be the case for the zero band gap, nearly charge neutral surface states as the electrons and holes are readily excited thermally. Based on the Boltzmann transport equation, the electronic thermal conductivity $k_e$ can be written as[34]

$$k_e = k_n + k_p + \frac{\sigma_n \sigma_p}{\sigma_n + \sigma_p}(\alpha_n - \alpha_p)^2 T \tag{3}$$

where $\alpha$ is the Seebeck coefficient. The third term comes from the bipolar diffusion which takes a maximum value when $\sigma_n = \sigma_p$, and it could be much larger than either $k_n$ and $k_p$. From the above equation, it is easy to tell that the bipolar contribution is in favor of small band gap (for high $\sigma_n$ and $\sigma_p$) as well as the Fermi level pinned around neutrality point to achieve $\sigma_n = \sigma_p$. While enhancing heat transfer, bipolar diffusion does not increase electrical conductivity (meaning $\sigma = \sigma_n + \sigma_p$), hence resulting in an increased Lorenz number. Experimentally, enhanced electronic thermal conductivity was previously observed at elevated temperature in small-gap semiconductors such as $Bi_2Te_3$[28], Al-based quasicrystals[37], and $In_4Se_{3-x}Te_x$[38], which was attributed to bipolar effect due to thermally excited electrons and holes. Further first-principles calculations predict that a large increase of Lorenz number (up to 10 times the Sommerfeld value) can occur in $Bi_2Te_3$[27,39] and skutterudites[40] when the Fermi level sits near the charge neutral point (center of the bulk band gap). Additionally, it has been predicted that the bipolar effect exists for Dirac electrons in graphene with a 1.8- to 4-fold increase of $L$ at room temperature, or any systems with a zero band gap and near charge neutrality condition[36]. Our BTS221 thin films do have a small band gap measured to be ~0.3 eV[23]. Moreover, as discussed above and in Supplementary Information, transport measurements indeed indicated that for thicknesses less than 20 nm, our samples were nearly charge neutral. Therefore, in our case, it is possible that the thermally populated surface Dirac electrons and holes, which are the major charge carriers at small thicknesses, are involved in the bipolar diffusion process that causes the observed high Lorenz number at the surface in our BTS221 thin films.

In summary, the thickness-dependent in-plane thermal conductivity of BTS221 TI thin films was measured and a large enhancement was observed as its thickness is reduced to less than 20 nm. Surface-dominated thermal and electrical transport was observed owing to the bulk-insulating nature of this material. Moreover, it is found that the Lorenz number for the TI surface is over 10 times larger than the Sommerfeld value. This is believed to stem from the unique transport properties of surface states, possibly due to the electrical and thermal current decoupling in the Dirac fluid as well as bipolar diffusion of electrons and holes at TI surface. Further work is needed to examine the relative importance of these two mechanisms (or possible involvement of additional mechanisms) in the observed enhancement of the surface Lorentz number.

METHODS

**Sample preparation.** BTS221 flakes were exfoliated using dicing tape onto poly(methyl methacrylate) (PMMA) and poly(vinyl alcohol) (PVA) thin film stack spin-coated onto Si wafer. The film stack with flakes was examined under the microscope to identify candidate flakes for transfer, then the desired flakes were precisely aligned to the holey silicon nitride membrane (Ted

Pella) and attached. The entire sample was then soaked into acetone to remove the PMMA, leaving the flakes suspended on the holes for Raman thermal measurements. The thin films were suspended on through-holes, which avoids additional laser absorption during Raman thermometry experiments. For electrical measurements, BTS221 flakes (typical thickness ~10–200 nm) are tape-exfoliated and placed on top of heavily doped Si substrates capped with 300 nm $SiO_2$. TI flakes with different thicknesses are selected under the optical microscope and characterized using atomic force microscopy (AFM). The four-terminal and Hall bar electrodes of the devices are fabricated by e-beam lithography, followed by e-beam deposition of Au/Cr (90/5 nm). The sample resistances $R$ are measured using the standard 4-terminal lock-in technique with an AC driving current of 1 μA at 13.33 Hz. The sample sheet resistance $R_S$ is calculated by $R_S = R \times W / L$, where $W$ and $L$ are the width and length of the conduction channel in the TI device, respectively. All the measurements are performed at room temperature and ambient pressure.

**Raman thermometry measurements.** A HORIBA LabRAM HR800 system was used. A 632.8 nm wavelength He-Ne laser focused by a 100x Olympus objective was used both as a heat source and Raman excitation source. The grating used has a groove density of 1800 l/mm. The Raman spectra were fitted using Lorentzian function to extract the Raman peak position. The reflectivity of each suspended film was measured by the same laser using a silver-coated mirror as reference with known reflectivity, while the transmissivity of the sample was obtained by a photon detector placed underneath it. Then the amount of absorbed laser power was derived from the reflectivity, transmissivity and the total incident laser power, and was subsequently used as heat source input for the numerical heat transfer model to extract the in-plane thermal conductivity of the suspended thin film.

REFERENCES

1. Moore, J. E. The birth of topological insulators. *Nature* **464,** 194–198 (2010).
2. Hasan, M. Z. & Kane, C. L. Colloquium: Topological insulators. *Rev. Mod. Phys.* **82,** 3045–3067 (2010).
3. Ando, Y. Topological insulator materials. *J. Phys. Soc. Japan* **82,** 102001 (2013).
4. Bernevig, B. A., Hughes, T. L. & Zhang, S.-C. Quantum spin Hall effect and topological phase transition in HgTe quantum wells. *Science* **314,** 1757–1761 (2006).
5. Hsieh, D. *et al.* A tunable topological insulator in the spin helical Dirac transport regime. *Nature* **460,** 1101–1105 (2009).
6. Qi, X.-L. & Zhang, S.-C. The quantum spin Hall effect and topological insulators. *Phys. Today* **63,** 33 (2010).
7. Chang, C.-Z. *et al.* Experimental Observation of the Quantum Anomalous Hall Effect in a Magnetic Topological Insulator. *Science* **340,** 167–170 (2013).
8. Xu, Y. *et al.* Observation of topological surface state quantum Hall effect in an intrinsic three-dimensional topological insulator. *Nat. Phys.* **10,** 956–963 (2014).
9. Jauregui, L. A., Pettes, M. T., Rokhinson, L. P., Shi, L. & Chen, Y. P. Magnetic field induced helical mode and topological transitions in a quasi-ballistic topological insulator nanoribbon with circumferentially quantized surface state sub-bands. *Nat. Nanotechnol.* **11,** 345–351 (2016).
10. Ren, Z., Taskin, A. A., Sasaki, S., Segawa, K. & Ando, Y. Large bulk resistivity and surface quantum oscillations in the topological insulator $Bi_2Te_2Se$. *Phys. Rev. B* **82,** 241306 (2010).
11. Taskin, A. A., Ren, Z., Sasaki, S., Segawa, K. & Ando, Y. Observation of dirac holes and electrons in a topological insulator. *Phys. Rev. Lett.* **107,** 16801 (2011).
12. Xiong, J. *et al.* High-field Shubnikov-de Haas oscillations in the topological insulator $Bi_2Te_2Se$. *Phys. Rev. B* **86,** 45314 (2012).
13. Ghaemi, P., Mong, R. S. K. & Moore, J. E. In-plane transport and enhanced thermoelectric performance in thin films of the topological insulators $Bi_2Te_3$ and $Bi_2Se_3$. *Phys. Rev. Lett.* **105,** 166603 (2010).

14. Luo, X., Sullivan, M. B. & Quek, S. Y. First-principles investigations of the atomic, electronic, and thermoelectric properties of equilibrium and strained $Bi_2Se_3$ and $Bi_2Te_3$ including van der Waals interactions. *Phys. Rev. B* **86,** 184111 (2012).
15. Xu, Y., Gan, Z. & Zhang, S.-C. Enhanced thermoelectric performance and anomalous seebeck effects in topological insulators. *Phys. Rev. Lett.* **112,** 226801 (2014).
16. Rittweger, F., Hinsche, N. F., Zahn, P. & Mertig, I. Signature of the topological surface state in the thermoelectric properties of $Bi_2Te_3$. *Phys. Rev. B* **89,** 35439 (2014).
17. Cai, W. *et al.* Thermal transport in suspended and supported monolayer graphene grown by chemical vapor deposition. *Nano Lett.* **10,** 1645–1651 (2010).
18. Yan, R. *et al.* Thermal Conductivity of Monolayer Molybdenum Disulfide Obtained from Temperature-Dependent Raman Spectroscopy. *ACS Nano* **8,** 986–993 (2014).
19. Zhou, H. *et al.* High thermal conductivity of suspended few-layer hexagonal boron nitride sheets. *Nano Res.* **7,** 1232–1240 (2014).
20. Peimyoo, N. *et al.* Thermal conductivity determination of suspended mono- and bilayer $WS_2$ by Raman spectroscopy. *Nano Res.* **8,** 1210–1221 (2015).
21. Luo, Z. *et al.* Measurement of In-Plane Thermal Conductivity of Ultrathin Films Using Micro-Raman Spectroscopy. *Nanoscale Microscale Thermophys. Eng.* **18,** 183–193 (2014).
22. Luo, Z. *et al.* Anisotropic in-plane thermal conductivity observed in few-layer black phosphorus. *Nat. Commun.* **6,** 8572 (2015).
23. Cao, H. *et al.* Controlling and distinguishing electronic transport of topological and trivial surface states in a topological insulator. *arXiv Prepr. arXiv1409.3217* 1–27 (2014). at <http://arxiv.org/abs/1409.3217>
24. Pan, Z. H. *et al.* Measurement of an exceptionally weak electron-phonon coupling on the surface of the topological insulator $Bi_2Se_3$ using angle-resolved photoemission spectroscopy. *Phys. Rev. Lett.* **108,** 187001 (2012).
25. Zhu, X. *et al.* Interaction of phonons and Dirac fermions on the surface of $Bi_2Se_3$: A strong Kohn anomaly. *Phys. Rev. Lett.* **107,** 186102 (2011).
26. Qiu, B., Sun, L. & Ruan, X. Lattice thermal conductivity reduction in $Bi_2Te_3$ quantum wires with smooth and rough surfaces: A molecular dynamics study. *Phys. Rev. B* **83,** 35312 (2011).

27. Pettes, M. T., Maassen, J., Jo, I., Lundstrom, M. S. & Shi, L. Effects of surface band bending and scattering on thermoelectric transport in suspended bismuth telluride nanoplates. *Nano Lett.* **13,** 5316–5322 (2013).

28. Goldsmid, H. The thermal conductivity of bismuth telluride. *Proc. Phys. Soc. B* **69,** 203–209 (1956).

29. Chang, J., Jadaun, P., Register, L. F., Banerjee, S. K. & Sahu, B. Dielectric capping effects on binary and ternary topological insulator surface states. *Phys. Rev. B* **84,** 155105 (2011).

30. Jenkins, G. S. *et al.* Dirac cone shift of a passivated topological $Bi_2Se_3$ interface state. *Phys. Rev. B* **87,** 155126 (2013).

31. Crossno, J. *et al.* Observation of the dirac fluid and the breakdown of the wiedemann-franz law in graphene. *Science* **351,** 1058 (2016).

32. Foster, M. S. & Aleiner, I. L. Slow imbalance relaxation and thermoelectric transport in graphene. *Phys. Rev. B* **79,** 85415 (2009).

33. Müller, M., Fritz, L. & Sachdev, S. Quantum-critical relativistic magnetotransport in graphene. *Phys. Rev. B* **78,** 115406 (2008).

34. Goldsmid, H. J. *Introduction to thermoelectricity*. (Springer, 2010). doi:10.1007/978-3-642-00716-3

35. Weathers, A. *et al.* Significant electronic thermal transport in the conducting polymer poly(3,4-ethylenedioxythiophene). *Adv. Mater.* **27,** 2101–2106 (2015).

36. Yoshino, H. & Murata, K. Significant enhancement of electronic thermal conductivity of two-dimensional zero-gap systems by bipolar-diffusion effect. *J. Phys. Soc. Japan* **84,** 24601 (2015).

37. Takeuchi, T. Thermal conductivity of the Al-based quasicrystals and approximants. *Z. Krist.* **224,** 35–38 (2009).

38. Rhyee, J. S., Cho, E., Ahn, K., Lee, K. H. & Lee, S. M. Thermoelectric properties of bipolar diffusion effect on $In_4Se_{3-x}Te_x$ compounds. *Appl. Phys. Lett.* **97,** 152104 (2010).

39. Huang, B.-L. & Kaviany, M. Ab initio and molecular dynamics predictions for electron and phonon transport in bismuth telluride. *Phys. Rev. B* **77,** 125209 (2008).

40. Chaput, L., Pécheur, P., Tobola, J. & Scherrer, H. Transport in doped skutterudites: Ab initio electronic structure calculations. *Phys. Rev. B* **72,** 85126 (2005).

ACKNOWLEDGEMENTS

This work was partly supported by DARPA MESO (grant number N66001-11-1-4107) and NSF (EFMA-1641101). The authors also thank Dr. Joseph Heremans for valuable discussions.

AUTHOR CONTRIBUTIONS

X.X. and Y.P.C. conceived the idea and supervised the experiments. Z.L. led the thermal conductivity experiments, Z.L. and M.S. performed the thermal conductivity measurements, and Z.L. analyzed the thermal conductivity data. J.T. performed and analyzed the electrical transport measurements. Z.L. J.T. and X.X. co-wrote the manuscript with input from all authors.

COMPETING FINANCIAL INTERESTS STATEMENT

The authors declare no competing financial interests.

# Supplementary Information for:

# Large Enhancement of Electronic Thermal Conductivity and Lorenz Number in Topological Insulator $Bi_2Te_2Se$ Thin Films

Zhe Luo[1,#], Jifa Tian[2,#], Mithun Srinivasan[1], Yong P. Chen[2,3,]* and Xianfan Xu[1,]*

[1] *School of Mechanical Engineering and Birck Nanotechnology Center, Purdue University, West Lafayette, Indiana 47907, USA*

[2] *Department of Physics & Astronomy and Birck Nanotechnology Center, Purdue University, West Lafayette, Indiana 47907, USA*

[3] *School of Electrical and Computer Engineering and Purdue Quantum Center, Purdue University, West Lafayette, Indiana 47907, USA*

[#] These authors contributed equally to this work.

** To whom correspondence should be addressed. Email: yongchen@purdue.edu, xxu@ecn.purdue.edu

**Supplementary Note 1 | Raman thermometry**

In Raman thermal measurements, the He-Ne laser was focused using 100x objective lens at the center of the suspended BTS films, creating a steady-state heat source. The Raman spectra collected were used to determine the local temperature at the laser focal spot, by examining the Raman peak red-shift. The temperature coefficients are used to convert the Raman peak shift to temperature. The Raman-measured temperature ($\theta_{Raman}$) is a Gaussian-weighted average temperature within the laser focal spot

$$\theta_{\text{Raman}} = \frac{\int_0^\infty \theta(r) \exp\left(-\frac{r^2}{r_0^2}\right) r \, dr}{\int_0^\infty \exp\left(-\frac{r^2}{r_0^2}\right) r \, dr}$$

With laser power changing, $d\theta_{\text{Raman}}/dP$ can be calculated and used to extract the in-plane thermal conductivity using a numerical heat transfer model.

To characterize the laser heat source, the optical reflectivity and transmissivity are measured on thin films exhibiting transparency at He-Ne wavelength 632.8 nm. A Newport 1815-C power meter is placed under the sample to measure the optical transmissivity $T$. A beam splitter is placed in the incident laser beam path to direct the reflected light from the sample into the power meter. By replacing the sample with a silver mirror as a reference, the reflectivity $R$ of the sample can be derived. The uncertainty of $R$ and $T$ is generally within 0.5% in absolute value. With $R$ and $T$ determined, the optical absorptivity $A$ can be calculated by $A = 1 - R - T$, and the absorption coefficient $\alpha$ was calculated by

$$\alpha = -\frac{1}{t} \ln\left(\frac{T}{1-R}\right)$$

The obtained value for our BTS films is $\alpha = 4.63 \pm 0.50 \times 10^7$ m$^{-1}$.

**Supplementary Note 2 | Uncertainty of the Lorenz number**

The uncertainty of $L_{\text{TSS}}$ is determined using the standard error propagation method, which is the process of calculating the uncertainty of a quantity that is derived from other parameters whose uncertainties are already available. In this case, $L_{\text{TSS}}$ value is derived from $\Gamma_{\text{TSS}}$ and $G_{\text{TSS}}$, and their uncertainties are obtained from the linear fitting. Employing the error propagation method, the uncertainty of $L_{\text{TSS}}$ can be determined as follows

$$\Delta L_{\text{TSS}} = \sqrt{\left(\frac{\partial L_{\text{TSS}}}{\partial \Gamma_{\text{TSS}}}\right)^2 \Delta \Gamma_{\text{TSS}}^2 + \left(\frac{\partial L_{\text{TSS}}}{\partial G_{\text{TSS}}}\right)^2 \Delta G_{\text{TSS}}^2}$$

$$= \sqrt{\left(\frac{1}{TG_{\text{TSS}}}\right)^2 \Delta \Gamma_{\text{TSS}}^2 + \left(-\frac{\Gamma_{\text{TSS}}}{TG_{\text{TSS}}^2}\right)^2 \Delta G_{\text{TSS}}^2}$$

$$= \sqrt{\left(\frac{L_{\text{TSS}}}{\Gamma_{\text{TSS}}}\right)^2 \Delta \Gamma_{\text{TSS}}^2 + \left(\frac{L_{\text{TSS}}}{G_{\text{TSS}}}\right)^2 \Delta G_{\text{TSS}}^2}$$

and the uncertainty value is reported in the main text.

**Supplementary Note 3 | Suspended four-terminal BTS device for validation purpose**

On a 200-nm-thick SiN membrane (Ted Pella), FEI Quanta focused-ion-beam was used to cut a ~3 μm diameter hole and to deposit ~70-nm-thick Pt contacts around the hole, as shown in Supplementary Fig. 1. A 15-nm-thick $Bi_2Te_2Se$ flake was transferred on to the structure via the PMMA/PVA transfer technique described in the main text. Electrical and thermal measurements were conducted on the same sample, and the results are shown as black pyramid symbols in the main text Fig. 3a and d, which agrees well with the other experimental data as well as the theoretical prediction.

**Supplementary Note 4 | Thickness dependence of the electrical field and Hall effects of $Bi_2Te_2Se$ samples.**

We have systematically studied the electrical field and Hall effects of the $Bi_2Te_2Se$ samples at various thicknesses. A representative Hall bar device with a thickness of 16 nm is shown in Supplementary Fig. 2. Supplementary Fig. 3 represents the corresponding sheet resistance $R_s$ as a function of back gate voltage ($V_g$) and Hall resistance as a function of magnetic field (B) measured at room temperature. We also extract the corresponding carrier density from the Hall resistance.

We find that the thick $Bi_2Te_2Se$ sample (> 30 nm) always shows relatively week field effect (Supplementary Figs 3a,c) and has a small Hall coefficient (Supplementary Figs 3b,d). From the Hall coefficient of the thick sample, we can determine that the charge carrier type is hole and the extracted 2D carrier density is on the order of $10^{14}$ cm$^{-2}$ (Supplementary Figs 3b,d), suggesting that the corresponding Fermi level is in the valence band and the transport is dominated by bulk carriers. As the thickness decreases, the field effect is notably enhanced and the carrier density is significantly decreased. When the sample thickness is below 20 nm (Supplementary Figs 3g–j), we find that the charge neutrality point (Supplementary Figs 3g,i) is much closer to zero gate voltage ($V_g$) than those of thick samples, indicating the transport is dominated by TSS and the corresponding Fermi level is very close to the Dirac point of TSS at room temperature. We further find that the sample Hall coefficient changes sign as the sample thickness decreases, indicating the carrier type changes from hole to electron with a low carrier density. Furthermore, the non-linear Hall resistance observed in the 11 nm-thick sample suggests the coexistence of the electrons and holes, where the real carrier density will be much lower than the "single-band" carrier density extracted from the apparent Hall coefficient ($1.16 \times 10^{13}$ cm$^{-2}$). Our demonstration of the Fermi level close to Dirac point of TSS and the non-linear Hall resistance in the very thin samples support the bipolar carrier transport picture we proposed in the main text.

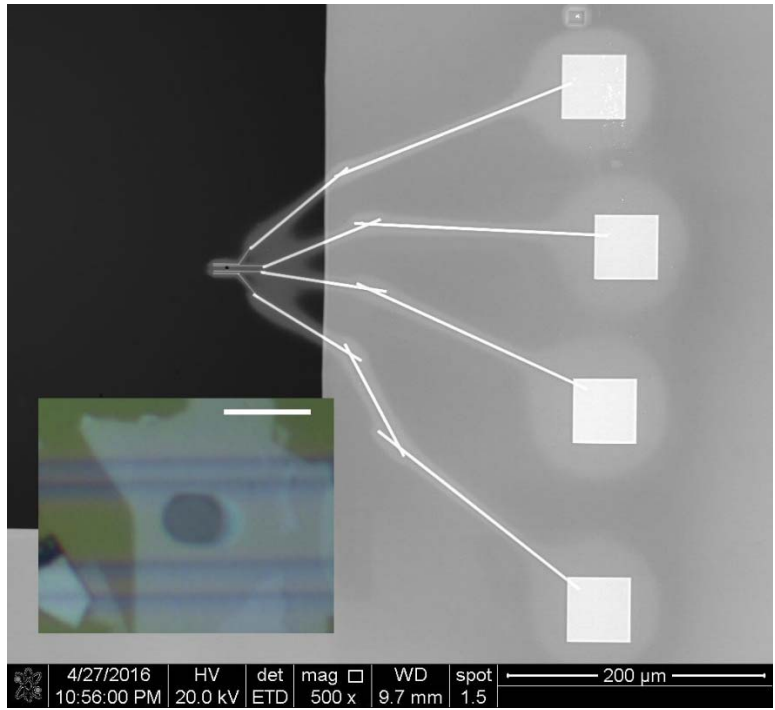

**Supplementary Figure 1 | SEM image of Pt contacts and optical image of suspended Bi$_2$Te$_2$Se device on SiN membrane (inset).** Scale bar of the inset is 5 μm.

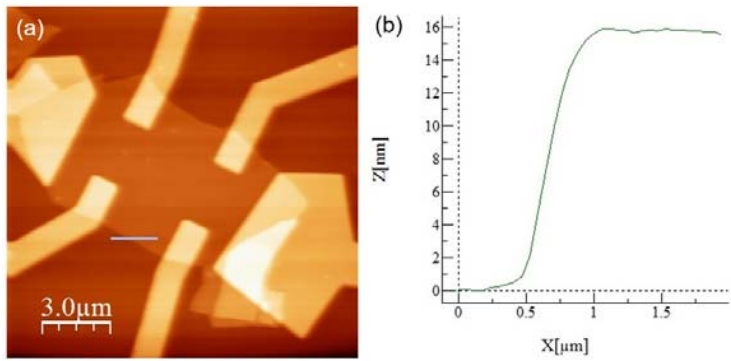

**Supplementary Figure 2 | A 16 nm-thick Bi$_2$Te$_2$Se device with Hall bar structures.** (a) AFM image; (b) the corresponding height profile of this device.

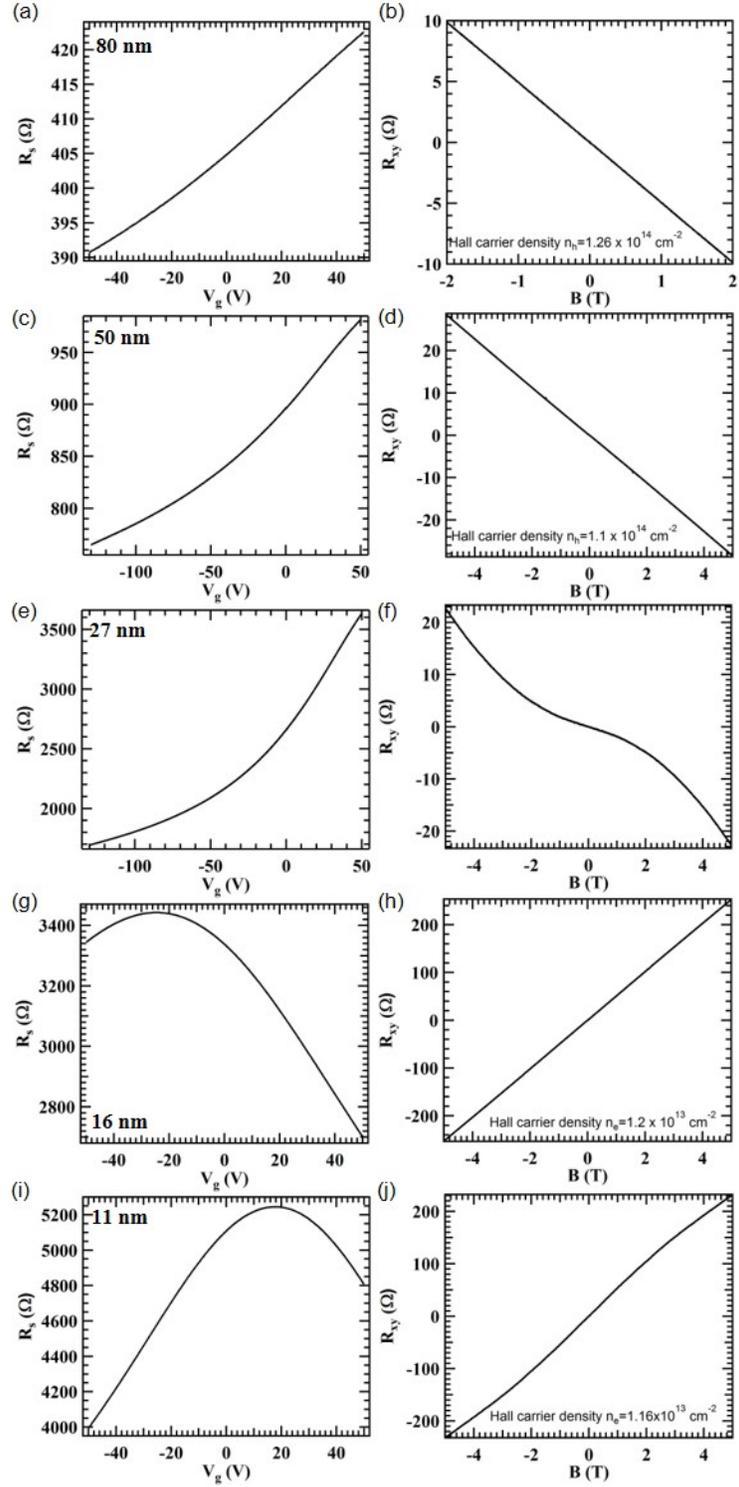

**Supplementary Figure 3 | Field and Hall measurement data.** The room temperature electrical field (a, c, e, g and i) effect and Hall (b, d, f, h, and j) effect of $Bi_2Te_2Se$ samples with various thicknesses of (a,b) 80 nm, (c,d) 50 nm, (e,f) 27 nm, (g,h) 16 nm, and (i,j) 11 nm.